\documentclass[11pt]{article}
\usepackage{amsfonts}
\usepackage{amsmath}
\usepackage{amssymb}
\usepackage{amsthm}
\usepackage{mathrsfs}
\usepackage{psfrag}
\usepackage{color}

\usepackage{graphicx}

\theoremstyle{plain}

\theoremstyle{definition}

\numberwithin{exercise}{section}
\numberwithin{equation}{section}
\numberwithin{theorem}{section}
\numberwithin{problem}{section}

\numberwithin{figure}{section}

\newcommand{\D}{\,\mathrm{d}}

\begin{document}

\title{{A commentary on the mathematical model of the heterogeneous gypsy moth larva population by G. Dwyer and his co-authors}}

\author{Artem S. Novozhilov\footnote{artem.novozhilov@ndsu.edu} \\[3mm]
\textit{\normalsize Department of Mathematics, North Dakota State University, Fargo, ND 58108, USA}}

\date{}

\maketitle

\begin{abstract}
In this commentary I utilize the general methods of the mathematical theory of heterogeneous populations in order to point out an omission in the analysis of the mathematical model in [Dwyer et al (2000), Am Nat, 156(2):105--120], which led to the conclusion in [Elderd et al (2008) Am Nat, 172(6):829--842] that the original model must be replaced with an alternative one because of the new data. I show that more thorough understanding of the underlying model allows twitching the model parameters to account for the observed data.

\end{abstract}

\section{Introductory remarks}
In a series of influential papers \cite{Dwyer2002,Dwyer2000,Dwyer1997,elderd2008host} Greg Dwyer and his co-authors proposed a general mathematical model to account for host heterogeneity among individual gypsy moth larva with respect to susceptibility to the virus. These papers combined accurate mathematical modeling, laboratory dose-response experiments, field transmission experiments, and observations of naturally occurring populations. In the initial works it was concluded that ``a model incorporating host heterogeneity in susceptibility to the virus gives a much better fit to data on virus dynamics [...] than does a classical model''\cite{Dwyer1997} and ``our experimental estimates of virus transmission rates and levels of heterogeneity in susceptibility in gypsy moth populations give model dynamics that closely approximate the dynamics of real gypsy moth populations''\cite{Dwyer2000}. However, in a more recent work \cite{elderd2008host} the original model was replaced with an alternative one, because it was found that ``Our data show that heterogeneity in infection risk in this insect is so high that it leads to a stable equilibrium in the models, which is inconsistent with the outbreaks seen in North American gypsy moth populations''\cite{elderd2008host}. In more technical terms, it was observed that highly heterogeneous populations (for which the coefficient of variation $c$ is bigger than one) exhibit oscillations with large amplitudes, whereas the mathematical model can produce such behaviors only for $c<1$.

My first goal in this short note is to show that the conclusion to refute the initial mathematical model is based on a small omission in the original analysis. To wit, while the mathematical details presented in \cite{Dwyer2000,Dwyer1997} are mostly correct, they are based on the assumption that the host susceptibility is well approximated by the gamma distribution. Abstract methods of the theory of heterogeneous populations, as presented in \cite{Novozhilov2008,nov2009hetero,novozhilov2012} with applications to the disease dynamics or in \cite{karev2009,karev2010} with applications to general population and evolutionary models, yield a precise characterization of the possible dynamical regimes depending on the chosen initial probability distribution. In particular, they show that an assumption on the initial distribution of susceptibility to follow the gamma distribution severely restricts possible time-dependent model solutions. To support my claim I present a mathematical model of a highly heterogeneous population (with $c>1$) and stable oscillations. My second goal is to attract attention of the biological community to transparent and at the same time general mathematical models of heterogeneous populations, which contain the initial model by Dwyer et al as a particular case and whose underlying theory is very well worked out. Finally, the observations I discuss here should prompt to reevaluate the data in \cite{elderd2008host} or collect more data to select the mathematical model that better approximates available observation.

\section{Mathematical model}
In my exposition I will mostly follow the notation from \cite{Dwyer2000} with several minor changes. Let $s(t,\nu)$ be the density of susceptible hosts having the susceptibility that is characterized by the parameter value $\nu$. Therefore, $s(0,\nu)$ defines the initial distribution of susceptibility before the disease starts, and $S(t)=\int_\Omega s(t,\nu)\D \nu$ is the total density of the host population at time $t$. Let $P(t)$ be the density of infectious cadavers at $t$, $\tau$ be the time between infection and death, and $\mu$ be the breakdown rate of the cadavers on the foliage. Then the mathematical model takes the form
\begin{equation}\label{eq:1}
\begin{split}
  \frac{\partial s}{\partial t}(t,\nu) & =-\nu s(t,\nu)P(t), \\\
  \frac{\D P}{\D t}(t)  & =\int_\Omega \nu P(t-\tau)s(t-\tau,\nu)\D \nu-\mu P(t),
\end{split}
\end{equation}
where $\Omega$ is the set of admissible values of $\nu$, e.g., $\Omega=[0,\infty)$. The initial conditions are
\begin{equation}\label{eq:2}
\begin{split}
     s(0,\nu)&=s_0(\nu)=S_0p_0(\nu)=S_0p(0,\nu),\\
     P(0)&=P_0,
\end{split}
\end{equation}
where $S_0=S(0)$ is the total initial density of the host population, and $p_0(\nu)$ is the \textit{initial} distribution of the parameter $\nu$ in the host population (such that $\int_\Omega p_0(\nu)d\nu=1$ and $p_0(\nu)\geq 0$ when $\nu\in\Omega$). I will use the notation $p(t,\nu)$ for the \textit{current} susceptibility distribution, which is given  by
$$
p(t,\nu)=\frac{s(t,\nu)}{S(t)}=\frac{s(t,\nu)}{\int_{\Omega}s(t,\nu)\D \nu}\,.
$$
Inasmuch as of the most interest is the final epidemic size $x$ (i.e., the proportion of the population that gets infected during the epidemics), it is possible to allow the time to go to infinity to obtain a transcendental equation for $x$ in terms of the initial conditions and model parameters. Instead of this approach, that was used by Dwyer et al, I will show some intermediate steps.

First, integration of the first equation in \eqref{eq:1} with respect to $\nu$ implies
\begin{equation}\label{eq:2a}
\begin{split}
  \frac{\D S}{\D t}(t) & =-\overline{\nu}(t) S(t)P(t), \\
  \frac{\D P}{\D t}(t)  & =\overline{\nu}(t-\tau)S(t-\tau)P(t-\tau)-\mu P(t),
\end{split}
\end{equation}
where
$$
\overline{\nu}(t)=\int_\Omega \nu p(t,\nu)\D\nu,
$$
is the current \emph{mean} of the distribution of $\nu$ in the host population at the moment $t$. It is not constant, but a function of time (intuitively, it must decrease, since the infection washes out first those who have initially higher values of $\nu$).

In \cite{Novozhilov2008,novozhilov2012} it was shown that function $\overline{\nu}(t)$ can be actually found if the moment generation function $M(\lambda)$ of the initial distribution is known, recall that it is defined as
\begin{equation}\label{eq:3}
M(\lambda)=\int_\Omega p_0(\nu)e^{\lambda\nu}\D\nu.
\end{equation}
It turns out that
\begin{equation}\label{eq:4}
\overline{\nu}(t)=\frac{\D}{\D\lambda} \log M(\lambda)|_{\lambda=q(t)}\,,
\end{equation}
where $q(t)$ solves
\begin{equation}\label{eq:5}
\frac{\D q}{\D t}(t)=-P(t),
\end{equation}
with the initial condition $q(0)=0$. The full details of this derivation can be found in the references above. Therefore, instead of two equations in \eqref{eq:1} I end up with three ODE \eqref{eq:2a}, \eqref{eq:4}, \eqref{eq:5}, which are equivalent to formally infinite dimensional system \eqref{eq:1}. Furthermore, it can be shown that \eqref{eq:2a}, \eqref{eq:4}, \eqref{eq:5} are equivalent to
\begin{equation}\label{eq:6}
\begin{split}
  \frac{\D S}{\D t}(t) & =-h(S(t))P(t), \\
  \frac{\D P}{\D t}(t)  & =h(S(t-\tau))P(t-\tau)-\mu P(t),
\end{split}
\end{equation}
where function $h(S)$ is given by ($M^{-1}$ is the inverse function to $M$)
$$
h(S)=S_0\frac{\D}{\D\lambda}M^{-1}(\lambda)|_{\lambda=S(t)/S_0}\,.
$$
This is the main theoretical result of \cite{Novozhilov2008} that says, quite surprisingly, that the dynamics of a heterogeneous population can be described by the same number of ordinary differential equations. I remark that no special assumption was made so far about the initial susceptibility distribution.

Now assume that the initial distribution is a gamma distribution with parameters $m$ and $k$:
$$
p_0(\nu)=\frac{m^k}{\Gamma(k)}\nu^{k-1}e^{-\nu m}\,.
$$
It has the mean and the variance
$$
\textrm{E}(X)=\frac{k}{m}\,,\quad \textrm{Var}(X)=\frac{k}{m^2}\,.
$$
Its moment generating function is
$$
M(\lambda)=\left(1-\frac{\lambda}{m}\right)^{-k}\,,
$$
and therefore, using the theory outlined above, I can find that for the initial gamma distribution system \eqref{eq:1} takes the form
\begin{equation}\label{eq:7}
\begin{split}
  \frac{\D S}{\D t}(t) & =-\frac{k}{m}\left(\frac{S(t)}{S_0}\right)^{1/k}S(t)P(t), \\
  \frac{\D P}{\D t}(t)  & =\frac{k}{m}\left(\frac{S(t-\tau)}{S_0}\right)^{1/k}S(t-\tau)P(t-\tau)-\mu P(t),
\end{split}
\end{equation}
which coincides with equations (3)-(4) in \cite{elderd2008host}. To reiterate: system (3)-(4) in \cite{elderd2008host} is equivalent to the original system \eqref{eq:1} if and only if the initial distribution of $\nu$ is a gamma distribution. In \cite{Dwyer2000} it was stated that ``Our results, however, can also be derived without this assumption'', where by ``assumption'' the initial gamma distributions is meant. This is the point, where an incomplete mathematical analysis led to an incorrect conclusion, which was based on the approximation that the coefficient of variation, defined as the fraction of the standard deviation to the mean, is held constant.

Let me explain in more details. When populations evolve with time, their characteristics also evolve (the mean, variance, coefficient of variation are all changing in general), but the distribution itself stays often (not always) the same. For example, if the initial distribution of susceptibility is the gamma distribution, than the density $p(t,\nu)$, defined above, is again the density of the gamma distribution, but now with the parameters (see \cite{karev2009} for mathematical details)
$$
\textrm{E}(X(t))=\frac{k}{m-q(t)}\,,\quad \textrm{Var}(X(t))=\frac{k}{(m-q(t))^2}\,.
$$
Note that both the mean and variance decrease with time since $q(t)\leq 0$ and decreasing. Using these expressions, we see that $c=\sqrt{\textrm{Var}(X(t))}/\textrm{E}(X(t))$ remains constant in time. Therefore, the assumption to fix $c$ in \cite{Dwyer2000} was \textit{equivalent} to assuming that the initial distribution is the gamma distribution. It can be proved that the gamma distribution is the only distribution for which the coefficient of variation does not change with time in this sort of mathematical models.

\section{Toy example}
Let me here give an example of a population with the \textit{initial} coefficient of variation $c>1$, which exhibit stable oscillations.

For the following I will need the equation for the final epidemic size $x$
$$
1-x=M\left(-\frac{S_0x+P_0}{\mu}\right),
$$
which can be obtained from \eqref{eq:7}.

I consider a very general family of distributions, defined by its moment generating function
$$
M(\lambda)=\exp\left[-\rho\left(1-\left(\frac{\nu}{\nu-\lambda}\right)^m\right)\right]
$$
with parameters $\nu>0,\,m>-1,m\rho>0$. This is the co-called variance function distributions \cite{aalen2008survival}. I find that
$$
\textrm{E}(X(t))=\rho\left(\frac{\nu}{\nu-q(t)}\right)^m\frac{m}{\nu-q(t)}\,,\quad \textrm{Var}(X(t))=\textrm{E}(X(t))\frac{m+1}{\nu-q(t)}\,.
$$
Therefore, to guarantee that the coefficient of variation decreases with time, I should take $m<0,\,\rho<0$. For example, parameters $\rho = -3,\, m = -0.2,\nu = 0.6$ imply that $\textrm{E}(X(0))=1,\,c(0)=1.1547>1$, and both the mean and the coefficient of variation will decrease with time.

\begin{figure}[!ht]
\centering
\includegraphics[width=0.45\textwidth]{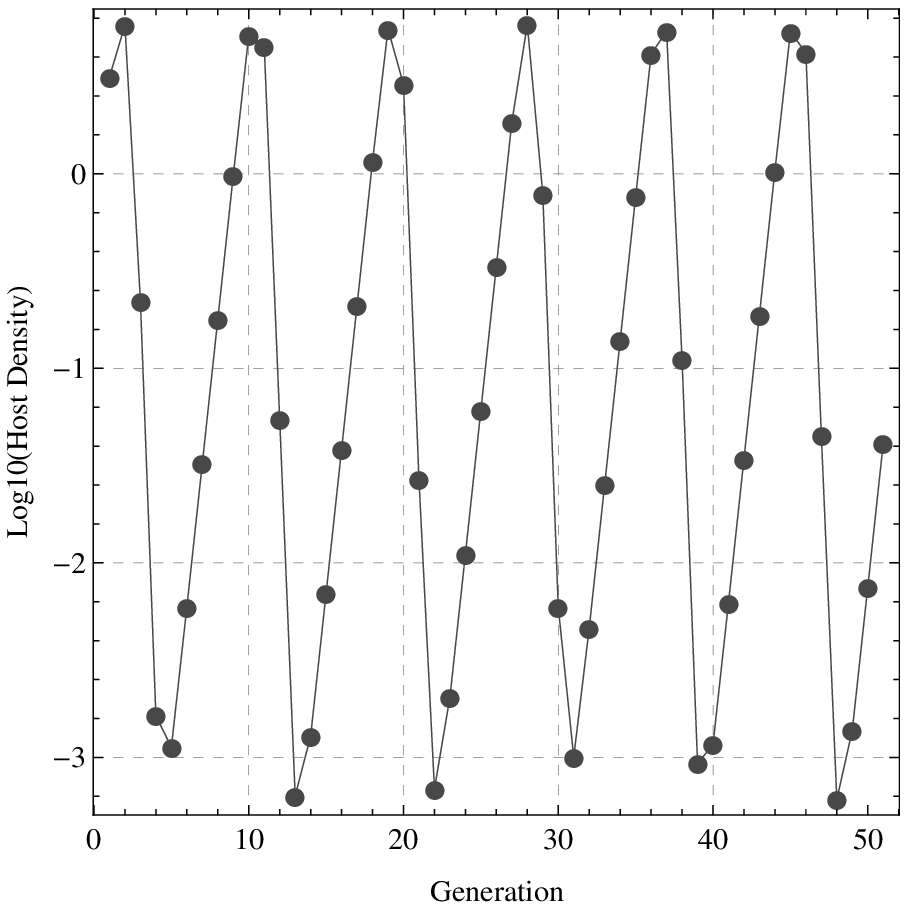}\hfill
\includegraphics[width=0.45\textwidth]{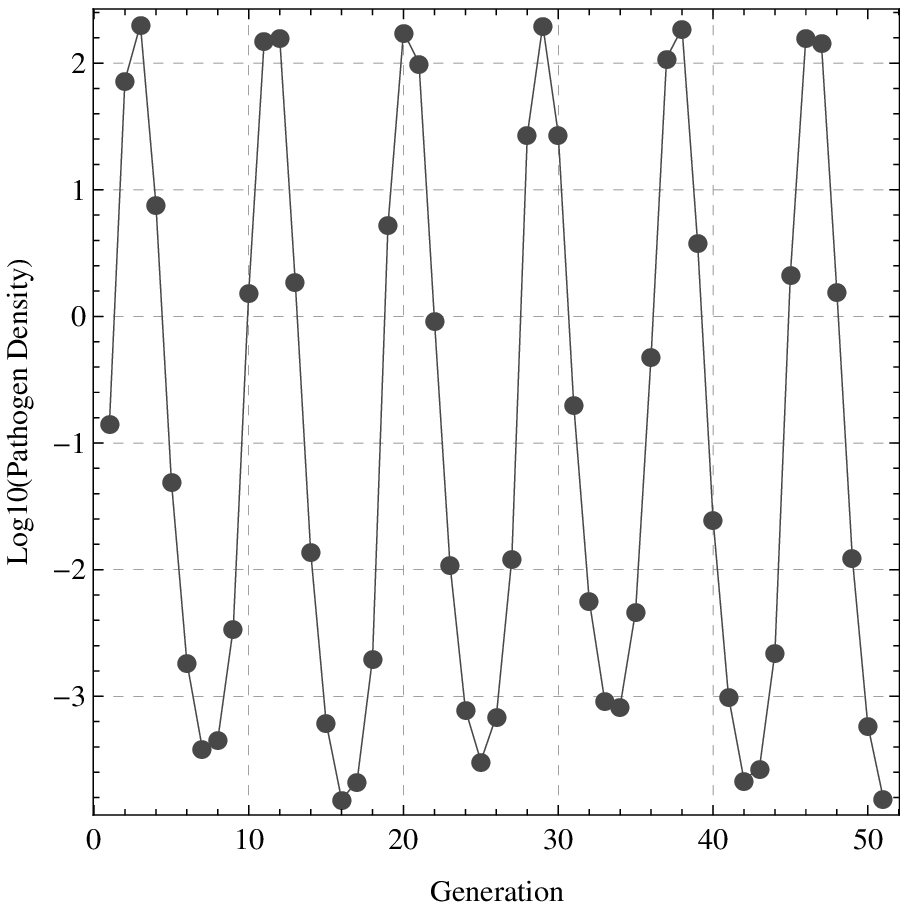}
\caption{Dynamics of the insect-pathogen models with the initial coefficient of variation exceeding 1. For the details and parameter values see the text}\label{fig:1}
\end{figure}

To produce population dynamics, I, following \cite{elderd2008host}, consider the system
\begin{align*}
N_{t+1}&=\lambda N_t(1-x_t),\\
Z_{t+1}&=\phi N_t x_t+\gamma Z_t,
\end{align*}
where $N_t$ and $Z_t$ are the initial host and pathogen densities in generation $t$, such that every time in the final epidemic size equation I take $S_0=N_t,\,P_0=Z_t$, $x_t$ is the final epidemic size for the year $t$, $\lambda,\phi,\gamma$ are the model specific parameters, which I take $\lambda=5.5,\phi=35,\gamma=0$ to coincide with the values used to produce Fig. 1 in \cite{elderd2008host}. The result of my simulations is given in Fig. \ref{fig:1}.
I note that despite the fact that the initial coefficient of variation is above 1 (the population is highly heterogeneous), we still observe stable oscillations of large amplitude. This is in contrast with Fig. 1B in \cite{elderd2008host}, where a stable equilibrium for the model with $c>1$ is shown. Therefore, in general, the conclusion from \cite{elderd2008host} that the model with the constant infection risk cannot produce the sustained oscillations if $c>1$ should be replaced with the conclusion that the model with the constant infection risk \textit{and} the initial gamma distribution of susceptibility cannot produce the sustained oscillations if $c>1$.

\section{Concluding remarks}
First of all, I would like to emphasize that my observations do not invalidate the conclusions in \cite{elderd2008host}. The important point here is that the mathematical model, originally suggested in \cite{Dwyer1997} and analyzed in \cite{Dwyer2000}, is still capable of producing dynamical regimes, which were thought to be impossible to observe in it (see my toy example above). Moreover, the general theory of heterogeneous populations \cite{karev2009,Novozhilov2008,novozhilov2012} provides a convenient and flexible tool to accommodate very different dynamical behaviors.

A straightforward test to determine whether the assumption on the initial gamma distribution in \cite{elderd2008host} should be replaced with something else may be performed by collecting estimates of the population coefficient of variation at different time moments during each epidemics. If these values do not deviate with time significantly, then it would support the reasonings in \cite{elderd2008host}. If, however, it would be possible to convincingly see that the coefficient of variation declines as the epidemics proceeds, then it should prompt to reevaluate the role of the model with the constant infection risk, as originally formulated by Dwyer and his co-authors. Unfortunately, the available data do not  seem to allow to perform such test at the moment.
\paragraph{Acknowledgements:} I would like to thank Bret Elderd for a profitable discussion while preparing this commentary.

\end{document}